\documentclass{PoS}

\usepackage{bm}

\title{Unquenching Effects on the Coefficients of the L\"uscher-Weisz Action}

\ShortTitle{Unquenching effects on the coefficients of the L\"uscher-Weisz action}

\author{Zh. Hao$^a$, \speaker{G.M. von Hippel}$^b$\thanks{new address: DESY Zeuthen, Platanenallee 6, 15738 Zeuthen, Germany.},
        R.R. Horgan$^c$, Q.J. Mason$^c$, H.D. Trottier$^a$\\ \\
        \llap{$^a$} Simon Fraser University, Department of Physics, Burnaby, BC, V5A 1S6, Canada \\
        \llap{$^b$} Department of Physics, University of Regina,  Regina, SK, S4S 0A2, Canada \\
        \llap{$^c$} DAMTP, CMS, University of Cambridge, Cambridge CB3 0WA, U.K.}

\abstract{The effects of unquenching on the perturbative improvement
          coefficients in the Symanzik action are computed within the
          framework of L\"uscher-Weisz on-shell improvement. We find
          that the effects of quark loops are surprisingly large, and
          their omission may well explain the scaling violations
          observed in some unquenched studies.}

\FullConference{The XXV International Symposium on Lattice Field Theory\\
                July 29 - August 4 2007\\
                Regensburg, Germany}

\begin{document}


\section{Introduction}
\label{intro}

Recent progress in parallel computing, as well as theoretical advances
in the formulation of lattice gauge theories with fermions, have allowed
the widespread adoption of simulations using
dynamical light quarks, leading to a significant
reduction in systematic errors by removing the uncontrollable
error inherent in the quenched approximation.

The Fermilab Lattice, MILC and HPQCD collaborations have
an ambitious program which to date has made several high-precision
predictions from unquenched lattice QCD simulations
\cite{Davies:2003ik, Aubin:2004fs}.
In this, we rely on the Symanzik-improved staggered-quark
formalism, specifically the use of the asqtad
\cite{Orginos:1999cr}
action. While this approach requires the use of the fourth root of the
staggered quark determinant, all the available evidence
is consistent with the conclusion that the resulting theory is in
the same universality class as continuum QCD, as long as the chiral
limit is taken after the continuum limit
\cite{Sharpe:2006re}.

Recent studies of the heavy-quark potential in full QCD
\cite{Davies:private}
have shown an unexpected apparent increase in scaling violations compared
to the quenched approximation. A possible reason for this would be that
these scaling violations arise from the mismatch between the inclusion of sea
quark effects in the simulation and the omission of sea quark effects
in the improvement coefficients in the action. This mismatch would appear to
spoil the $\mathcal{O}(a^2)$ improvement at the level of
$\mathcal{O}(\alpha_s N_f a^2)$. While a systematic study of
$\mathcal{O}(\alpha_sa^2)$ effects is generally beyond the scope of
the current perturbative improvement programme,
it is still important to bring up-to-date the calculations 
of the L\"uscher-Weisz improved gluonic action
\cite{Luscher:1985wf,Snippe:1997ru}
to include the effects of dynamical quarks.
This is important also because the L\"uscher-Weisz improvement is currently
included in many unquenched simulations 
\cite{Orginos:1999cr}.
Since the lattice spacing scale is set by measurement of the
heavy-quark potential, there will be an induced
$\mathcal{O}(\alpha_s N_f a^2)$ artifact by omitting the corrections
due to unquenching.
While such errors are generally smaller than other systematic errors in
current state-of-the art studies, it is simple to remove them using the
result of the perturbative matching calculations outlined here. For
details, the reader is referred to our paper
\cite{Hao:2007iz}.
%


\section{On-shell improvement}
\label{osi}

The L\"uscher-Weisz action is given by
\cite{Luscher:1984xn}
\begin{equation}\label{eqn:lw_action}
S = \sum_x \Bigg\{
c_0 \sum_{\mu\not=\nu}\left<1-P_{\mu\nu}\right>
+2 c_1 \sum_{\mu\not=\nu}\left<1-R_{\mu\nu}\right>
+\frac{4}{3} c_2 \sum_{\mu\not=\nu\not=\rho}\left<1-T_{\mu\nu\rho}\right>
\Bigg\}\;,
\end{equation}
where $P$, $R$ and $T$ are the plaquette, rectangle and ``twisted''
parallelogram loops, respectively. The requirement of obtaining the
Yang-Mills action in the continuum limit imposes the constraint
\begin{equation}
c_0 + 8 c_1 + 8 c_2 = 1\;,
\end{equation}
which fixes $c_0$ given the other two coefficients. This leaves us
with $c_1$ and $c_2$ to be determined in
order to eliminate the $\mathcal{O}(a^2)$ lattice artifacts.

If we have two independent quantities $Q_1$ and $Q_2$ which, at each
order in perturbation theory, can be
expanded in powers of $(\mu a)$, where $\mu$ is some energy scale, as
\begin{equation}
Q_i = \bar{Q}_i + w_i (\mu a)^2 + d_{ij} c_j (\mu a)^2 +
\mathcal{O}\left((\mu a)^4\right)\;,
\end{equation}
then the $\mathcal{O}(a^2)$ matching condition reads
\begin{equation}
\label{eqn:impcond_generic}
d_{ij} c_j = -w_i\;.
\end{equation}
Since this equation is linear, we can decompose the $w_i$ into a
gluonic and a fermionic part as $w_i = w_i^\textrm{glue} + N_f
w_i^\textrm{quark}$ and obtain the same decomposition for the $c_i$;
thus, especially we do not need to repeat the quenched calculation
in order to obtain the $\mathcal{O}(N_f)$ contributions.\footnote{Although
doing so provides a useful check on our methods, and we have in fact
successfully reproduced the results of \cite{Snippe:1997ru}.}

At tree-level, there are no fermion loops to consider,
and hence the tree-level coefficients remain unchanged
compared to the quenched case
\cite{Luscher:1985wf}:
\begin{equation}
c_1 ~=~ -\frac{1}{12}\;, \;\;\;\;\; c_2 ~=~ 0.
\end{equation}


\section{Lattice perturbation theory on a twisted lattice}
\label{tina1}

In lattice perturbation theory, the link variables $U_\mu\in SU(N)$ are
expressed in terms of the gauge field $A_\mu\in su(N)$ as
\begin{equation}
U_\mu(x) =
\exp\left(g a A_\mu\left(x+\frac{1}{2}\hat{\mu}\right)\right)
\end{equation}
which, when expanded in powers of $g$, leads to a perturbative
expansion of the lattice action, from which the perturbative vertex
functions can be read off.

As in any perturbative formulation of a gauge theory, gauge fixing and
ghost terms appear in the Fadeev-Popov Lagrangian; an additional term
arises from the Haar measure on the gauge group. Here we will not
have to concern ourselves with these, since for our purpose we only need
to consider quark loops.

To handle the complicated form of the vertices and propagators in
lattice perturbation theory, we employ a number of automation methods
\cite{Hart:2004bd,Trottier:2003bw}
that are based on the seminal work of L\"uscher and Weisz
\cite{Luscher:1985wf}.
Three independent implementations by different authors have been
used in this work to ensure against programming errors.

We work on a four-dimensional Euclidean lattice of length $La$ in the
$x$ and $y$ directions and lengths $L_za,~L_ta$ in the $z$ and $t$ 
directions, respectively, where $a$ is the lattice spacing and $L,L_z,L_t$ are
even integers. In the following, we will employ twisted boundary conditions
in much the same way as in
\cite{Luscher:1985wf,Snippe:1997ru}.
The twisted boundary conditions we use for gluons and quarks are
applied to the $(x,y)$ directions and are given by ($\nu=x,y$)
\begin{eqnarray}
U_\mu(x+L\hat{\nu}) & = & \Omega_\nu U_\mu(x) \Omega_\nu^{-1}\;, \\
\Psi(x+L\hat{\nu}) & = & \Omega_\nu \Psi(x) \Omega_\nu^{-1}\;,
\end{eqnarray}
where the quark field $\Psi_{sc}(x)$ becomes a matrix in smell-colour
space
\cite{Parisi:1984cy}
by the introduction of a ``smell'' group SU($N_s$) with $N_s=N$ in
addition to the colour group SU($N$). We apply periodic boundary
conditions in the $(z,t)$ directions.

These boundary conditions lead to a change in the Fourier expansion of
the fields: in the twisted $(x,y)$ directions the momentum sums are
now over
\begin{equation}
p_\nu = m n_\nu,~~-\frac{NL}{2} < n_\nu \le \frac{NL}{2},~~\nu = (x,y)\;,
\end{equation}
where $m = \frac{2\pi}{N L}$.
The modes with ($n_x=n_y=0 \textrm{ mod } N$) are omitted from the sum
in the case of the gluons. The momentum sums for quark loops
need to be divided by $N$ to remove the redundant smell factor.

The twisted theory can be viewed as a two-dimensional field theory in
the $(z,t)$ plane by considering the modes in the twisted
directions as Kaluza-Klein modes. Denoting $\mathbf{n}=(n_x,n_y)$, the
stable particles in the $(z,t)$ continuum limit of this effective
theory are called the A mesons ($\mathbf{n}=(1,0)$ or
$\mathbf{n}=(0,1)$) with mass $m$ and the B mesons
($\mathbf{n}=(1,1)$) with mass $\sqrt{2}m$
\cite{Snippe:1997ru}.
%


\section{Small-mass expansions}
\label{matl1}

To extract the $\mathcal{O}(a^2)$ lattice artifacts, we first expand
some observable quantity $Q$ in powers of $ma$ at fixed $m_qa$:
\begin{equation}\label{eqn:fit_in_ma}
Q(ma,m_qa)=a^{(Q)}_0(m_qa) + a^{(Q)}_2(m_qa) (ma)^2 +
\mathcal{O}\left((ma)^4,(ma)^4\log(ma)\right)\label{eqn:Q}
\end{equation}
where the coefficients in the expansion are all functions of $m_qa$.
There is no term at $\mathcal{O}\left((ma)^2\log(ma)\right)$ since the
gluon action is improved at tree-level to $O(a^2)$
\cite{Snippe:1997ru}.
Although we ultimately wish to extrapolate to the chiral limit, 
we cannot set $m_qa=0$ straight away, since the
correct chiral limit is $m_qa \to 0,~ma \to 0,~m_q/m > C$, where
$m = \frac{2\pi}{NL}$ as before and $C$ is a constant determined by
the requirement that a Wick rotation can be performed without
encountering a pinch singularity. This requires us to consider a
double expansion in $m_qa,ma$ and carry out the extrapolation to
$m_qa=0$ for the coefficients in Eqn. (\ref{eqn:Q}).

To extrapolate to the chiral limit, $m_qa \to 0$, we will fit the
coefficients in the expansion for $Q$ in $ma$ to their most general 
expansion in $m_qa$ for small $m_qa$. 

For $a^{(Q)}_0(m_qa)$ we have
\begin{equation}\label{eqn:fit0_in_mqa}
a^{(Q)}_0(m_qa)~=~b^{(Q)}_{0,0}\log(m_qa) + a^{(Q)}_{0,0}\;.
\end{equation}
Since we expect a well-defined continuum limit, $a^{(Q)}_0(m_qa)$
cannot contain any negative powers of $m_qa$ but, depending on the
quantity $Q$, it may contain logarithms; $b^{(Q)}_{0,0}$ is the
anomalous dimension associated with $Q$, and can be determined
by a continuum calculation.

For $a^{(Q)}_2(m_qa)$ we find
\begin{equation}\label{eqn:fit2_in_mqa}
a^{(Q)}_2(m_qa)~=~\frac{a^{(Q)}_{2,-2}}{(m_qa)^2} + a^{(Q)}_{2,0}
+ \left(a^{(Q)}_{2,2} + b^{(Q)}_{2,2}\log(m_qa)\right)(m_qa)^2 +
   \mathcal{O}\left((m_qa)^4\right)\;.\nonumber
\end{equation}
After multiplication by $(ma)^2$ the $(m_qa)^{-2}$ contribution gives
rise to a continuum contribution to $Q$, and $a^{(Q)}_{2,-2}$ is
calculable in continuum perturbation theory. There can be no term in
$(m_qa)^{-2}\log(m_qa)$ since this would be a volume-dependent further
contribution to the anomalous dimension of $Q$, and there can be no
term in $\log(m_qa)$ since the action is tree-level $O(a^2)$ improved.

In the chiral limit $m_q\to 0$, the term $w_i$ that appears on the
right-hand side of Eqn.
(\ref{eqn:impcond_generic})
is $a^{(Q)}_{2,0}$, and it is this limit and this coefficient that we
will concern ourselves with hereafter.


\section{The A meson mass}
\label{mtolo1}

\begin{figure}
\begin{center}
\includegraphics[width=5cm,keepaspectratio=,clip=]{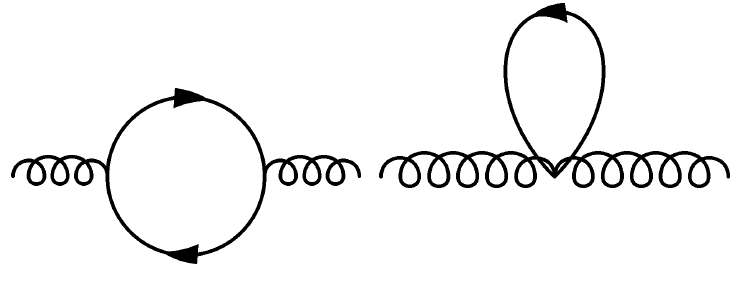}
\hspace{1cm}
\includegraphics[width=7cm,keepaspectratio=,clip=]{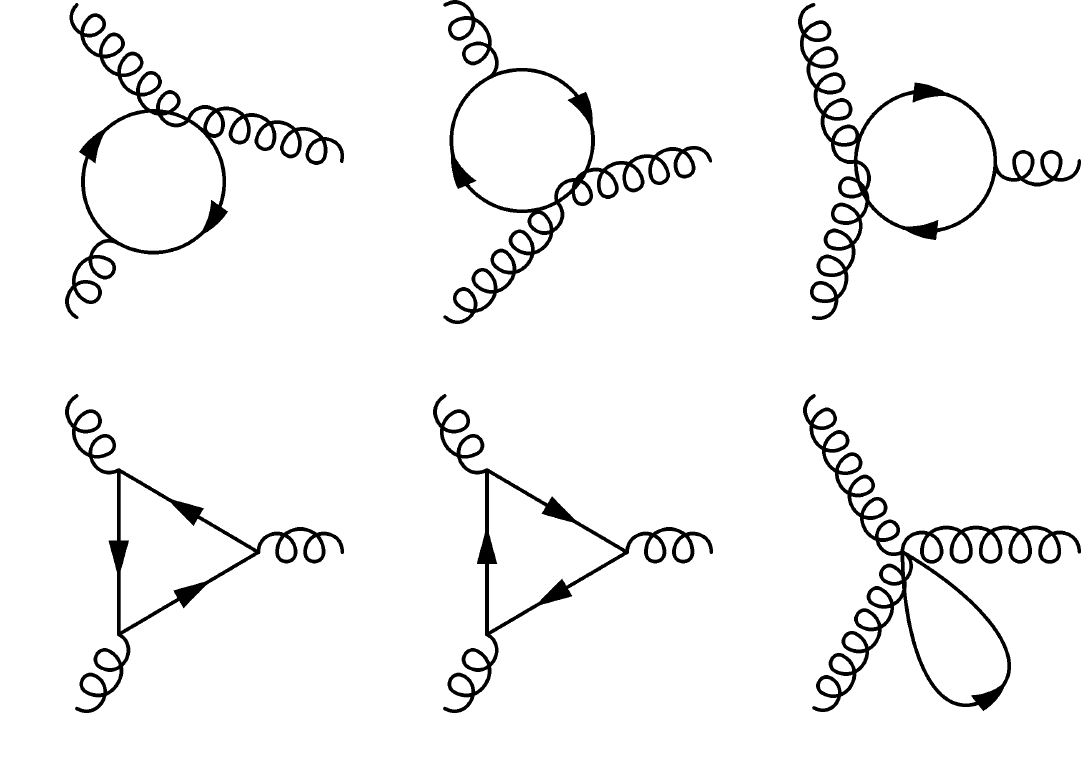}\\
\footnotesize (a)\hspace{7cm}(b)
\end{center}
\caption{(a) The fermionic one-loop diagrams contributing to the A meson
  mass renormalisation as well as to the wavefunction renormalisation
  for A and B mesons. (b) The fermionic diagrams contributing to the
  irreducible three-point function.}
\label{fig:diagrams}
\end{figure}

The simplest spectral quantity that can be chosen within the framework
of the twisted boundary conditions outlined above is the
(renormalised) mass of the A meson. The one-loop correction the the A
meson mass (for A mesons with positive spin) is given by
\begin{equation}\label{eqn:mA}
m_A^{(1)} = - Z_0(\mathbf{k})
              \left.\frac{\pi_{11}^{(1)}(k)}{2 m_A^{(0)}}\right|
              _{k=(i m_A^{(0)},0,m,0)}
\end{equation}
where $Z_0(\mathbf{k})=1+\mathcal{O}\left((ma)^4\right)$ is the
residue of the pole of the tree-level gluon propagator at spatial
momentum $\mathbf{k}$, and $m_A^{(0)}$ is defined so that the momentum
$k$ is on-shell. We consider the dimensionless quantity $m_A^{(1)}/m$.
The fermionic diagrams that contribute to this quantity are shown in
figure
\ref{fig:diagrams} (a).

The anomalous dimension of $m_A$ is zero and so using 
Eqn. (\ref{eqn:fit0_in_mqa}) we have $b^{(m_A,1)}_{0,0}~=~0$.
From gauge invariance we find $a^{(m_A,1)}_{2,-2}~=~0$
and $a^{(m_A,1)}_0(m_qa)=0$, which together with the previous result
implies that $a^{(m_A,1)}_{0,0}=0$.

The $\mathcal{O}\left(\alpha_s (ma)^2\right)$ contribution from
improvement of the action is given by
\cite{Snippe:1997ru}
\begin{equation}
\Delta_\textrm{imp} \frac{m_A^{(1)}}{m} = 
- ( c_1^{(1)} - c_2^{(1)} ) (ma)^2 + \mathcal{O}\left((ma)^4\right)\;.
\end{equation}


\section{The three-point coupling}
\label{tqff1}

An effective coupling constant $\lambda$ for an AAB meson vertex is
defined as
\begin{equation}
\label{eqn:def_of_lambda}
\lambda =
g_0 \sqrt{ Z(\mathbf{k}) Z(\mathbf{p}) Z(\mathbf{q}) }
e_j \Gamma^{1,2,j}(k,p,q)
\end{equation}
where we have factored out a twist factor of
$\frac{i}{N}\mathrm{Tr}([\Gamma_k,\Gamma_p]\Gamma_q)$ from both sides, and the
momenta and polarisations of the incoming particles are
\begin{equation}\label{eqn:momenta}
\begin{array}{llll}
k = (iE(\mathbf{k}),\mathbf{k}) \;\;\;\; &
p = (-iE(\mathbf{p}),\mathbf{p}) \;\;\;\;&
q = (0,\mathbf{q}) &  e = (0,1,-1,0) \\
\mathbf{k} = (0,m,ir)&\mathbf{p} = (m,0,ir)\;&\mathbf{q} = (-m,-m,-2ir)&
\end{array}
\end{equation}
Here $r>0$ is defined such that $E(\mathbf{q})=0$. This coupling is a
spectral quantity since it can be related to the scattering amplitude
of A mesons
\cite{Luscher:1985zq}.
We expand Eqn.
(\ref{eqn:def_of_lambda})
perturbatively to one-loop order and find (up to $\mathcal{O}((ma)^4)$
corrections)
\begin{equation}
\frac{\lambda^{(1)}}{m} =
\left( 1 - \frac{1}{24} m^2 \right)\frac{\Gamma^{(1)}}{m}
- \frac{4}{k_0} \frac{d}{dk_0}
           \left.\pi_{11}^{(1)}(k)\right|_{k_0=iE(\mathbf{k})}
 - \left( 1 - \frac{1}{12} m^2 \right) \frac{d^2}{dq_0^2}
           \left. \left( e^i e^j \pi_{ij}^{(1)}(q) \right) \right|_{q_0=0}
\end{equation}
The fermionic diagrams contributing to the irreducible three-point
function $\Gamma^{(1)}$ are shown in figure
\ref{fig:diagrams} (b).
Continuum calculations of the anomalous dimension and infrared
divergence give
\begin{equation}\label{eqn:anom_dir_lambda}
b^{(\lambda,1)}_{0,0}~=~-\frac{N_f}{3\pi^2}g^2\;,\;\;\;\;\;\
a^{(\lambda,1)}_{2,-2}~=~-\frac{N_f}{120\pi^2}g^2\;.
\end{equation} 

The improvement contribution to $\lambda$ is
\cite{Snippe:1997ru}
\begin{equation}
\Delta_\textrm{imp} \frac{\lambda^{1}}{m} = 4 (9 c_1^{(1)} - 7 c_2^{(1)}) (ma)^2
   + \mathcal{O}\left((ma)^4\right)\;.
\end{equation}


\section{Continuing to imaginary momenta}
\label{tnse1}

The external lines of the diagrams are on-shell,
but with complex three-momentum $\mathbf{k}$;
in the Euclidean formulation $k_0$ is also imaginary. In evaluating
the loop integrals that are not pure tadpoles, care must be taken to
ensure that the amplitudes calculated are the correct analytic
continuations from the Minkowski space on-shell amplitudes
defined with real three-momenta to the ones in
Eqn.
(\ref{eqn:momenta}).

The situation is complicated by the presence of two mass scales $m,
m_q$. The integrals are evaluated after performing a Wick rotation in
$k_0$, taking care to avoid contour crossing of any poles that move
as $r$ is continued from $r=0$ to $r=m/\sqrt{2}$. This requires
$m_q/m>C$, where $C$ is a constant dependent on the graph
being considered. After the Wick rotation in $k_0$, the (Euclidean)
integration contour for $k_0$ (or, in one case, $k_3$) must be
shifted by an imaginary constant.


\section{Results}
\label{tnvc1}

To extract the improvement coefficients from our diagrammatic
calculations, we compute the diagrams for a number of different values
of both $L$ and $m_q$ with $N_f=1$, $N=3$. At each value of $m_q$, we
then perform a fit in $ma$ of the form given in Eqn.
(\ref{eqn:fit_in_ma})
to extract the coefficients $a_n^{(Q,1)}(m_qa),~n=0,2$.

\begin{figure}
\includegraphics[height=4.8cm,angle=90,keepaspectratio=,clip=]{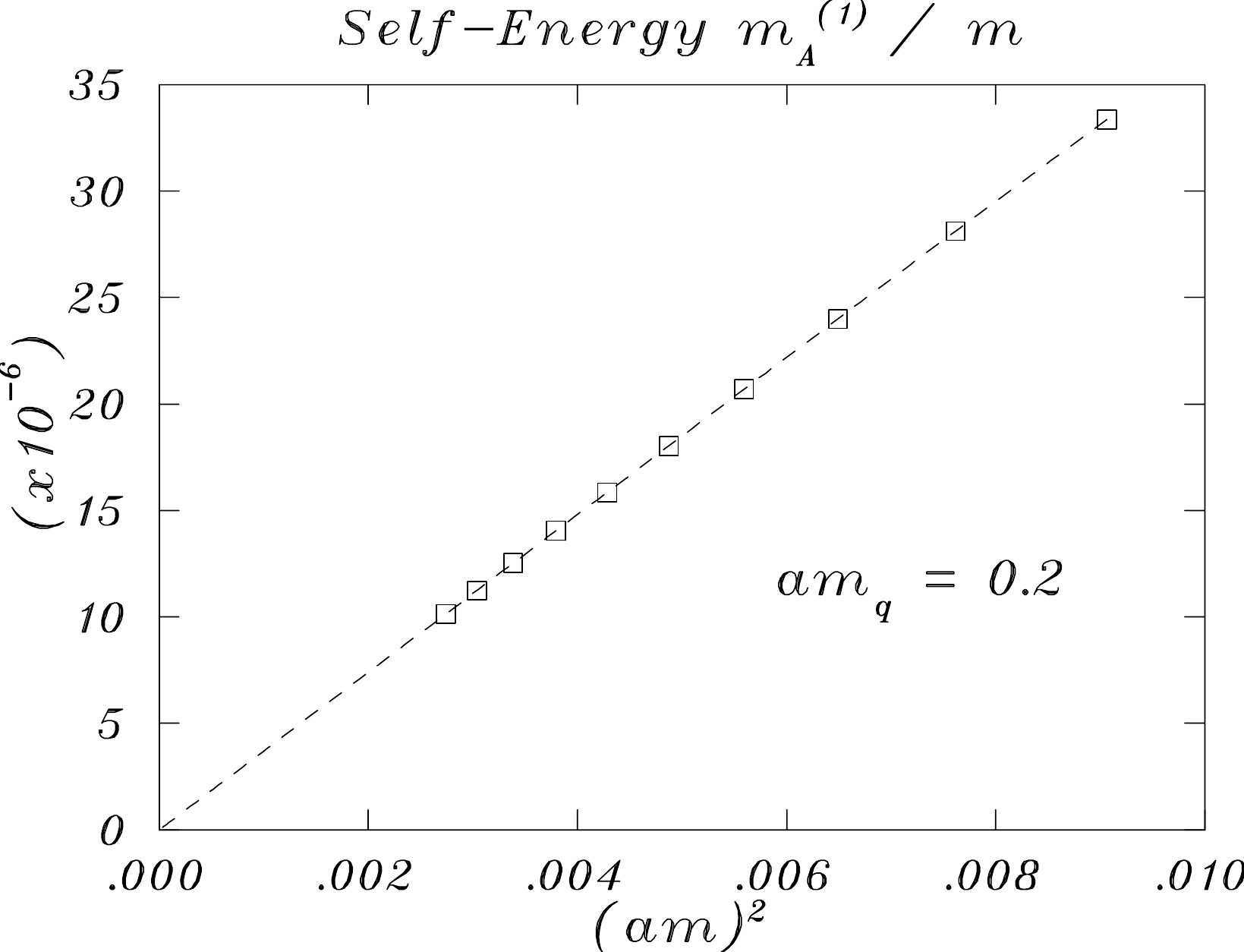}
\hfill
\includegraphics[height=4.8cm,angle=90,keepaspectratio=,clip=]{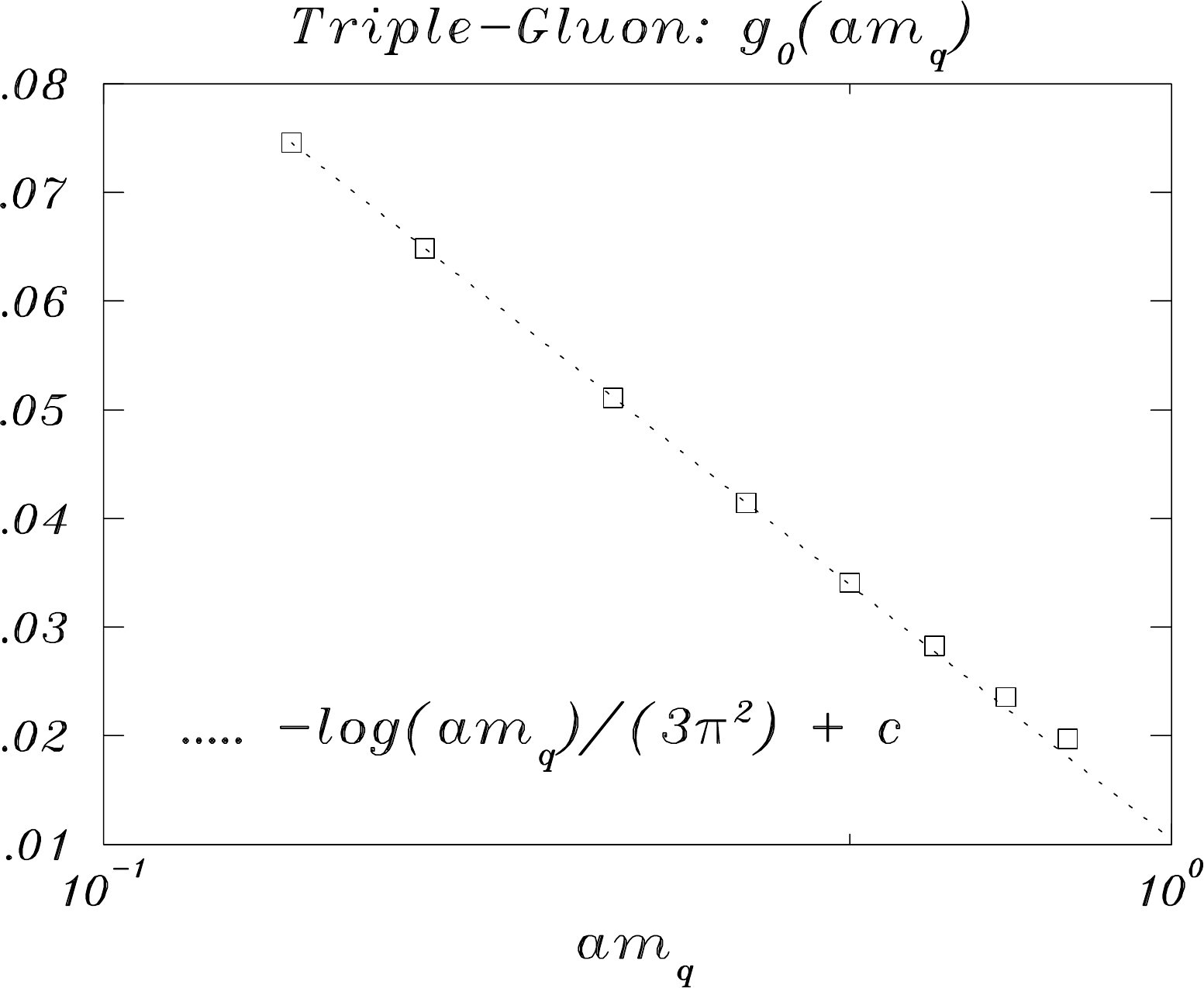}
\hfill
\includegraphics[height=4.8cm,angle=90,keepaspectratio=,clip=]{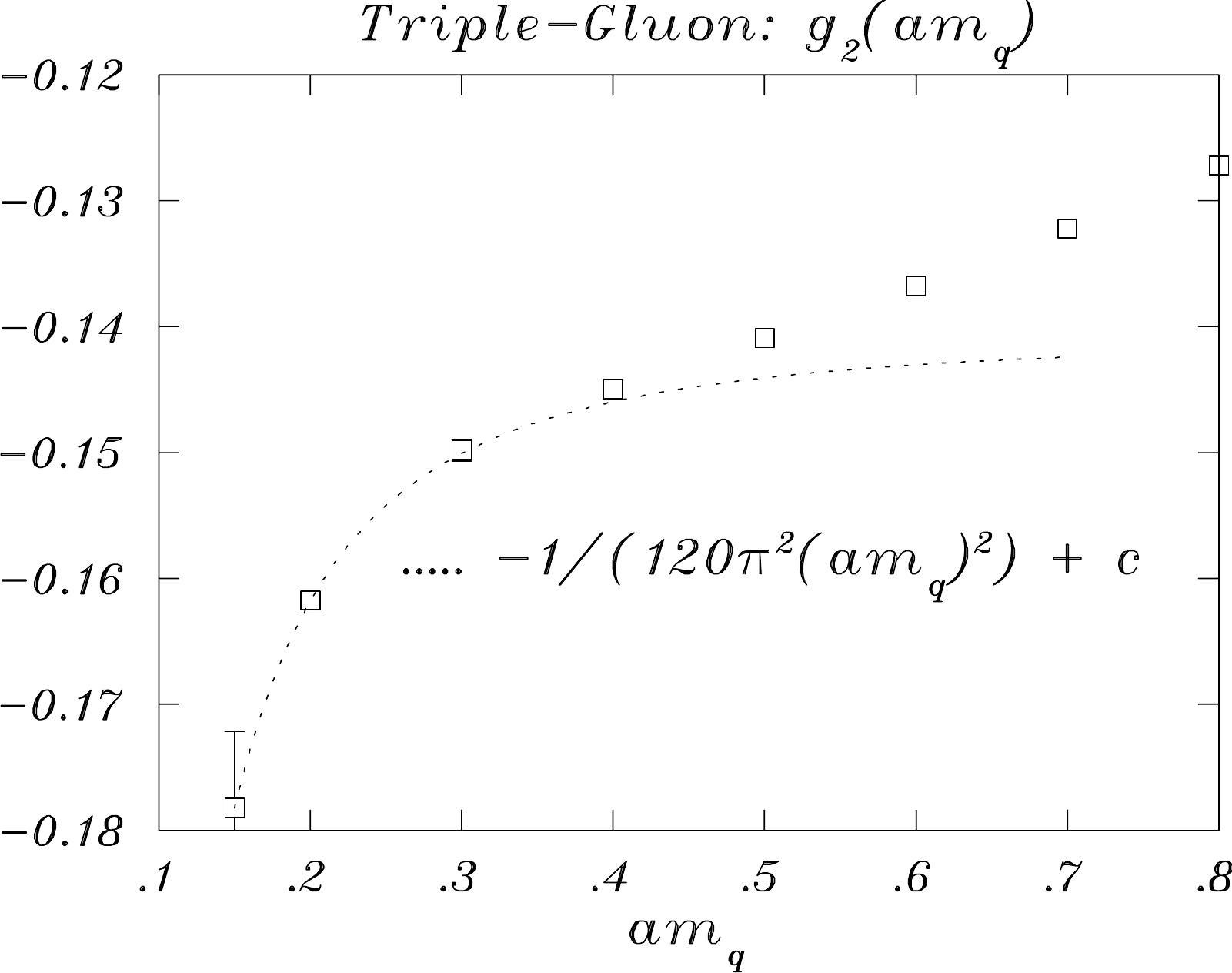}\\
\footnotesize${}$\hfill (a) \hfill\hfill (b) \hfill\hfill (c) \hfill${}$
\caption{(a) A plot of the fermionic contributions to the one-loop $A$
  meson self-energy $m_A^{(1)}/m$ against $(ma)^2$. The vanishing of
  $m_A^{(1)}/m$ in the infinite-volume limit can be seen clearly.
  (b) A plot of $a_0^{(\lambda,1)}$ against $m_qa$ which shows the
  agreement between the numerical lattice results and the known
  anomalous dimension.
  (c) A plot of $a_2^{(\lambda,1)}$ against $m_qa$ with the
  analytical continuum result for the infrared divergence shown for
  comparison.}
\label{fig:plots}
\end{figure}

Performing a fit of the form
(\ref{eqn:fit0_in_mqa}) and (\ref{eqn:fit2_in_mqa}),
respectively, on these coefficients, we get the required coefficients 
of the $\mathcal{O}(a^2)$ lattice artifacts in the chiral limit to be
\begin{eqnarray}
a_{2,0}^{(m_A,1)} & = & 0.00361(1) \\
a_{2,0}^{(\lambda,1)} & = & -0.140(1)
\end{eqnarray}
These coefficients are to be identified with the $w_i$ of Eqn.
(\ref{eqn:impcond_generic}).

Solving equation
(\ref{eqn:impcond_generic})
for $c_i^{(1)}$, our results can be summarised as
\begin{eqnarray}
c_1^{(1)} & = & -0.025218(4) + 0.00486(13) N_f \\
c_2^{(1)} & = & -0.004418(4) + 0.00126(13) N_f
\end{eqnarray}
where the quenched ($N_f=0$) results are taken from
\cite{Snippe:1997ru}.
With $N_f=3$ the shift from the quenched values is surprisingly large,
and may have a significant impact.


\end{document}